\documentclass[amsmath,twocolumn,amssymb,prl,preprintnumbers,showpacs]{revtex4}
\usepackage{amsmath}
\usepackage[cmtip,arrow]{xy}
\usepackage{pb-diagram,pb-xy}
\usepackage{slashed}
\usepackage{amssymb,amsopn}
\def \be {\begin{equation}}
\def \ee {\end{equation}}
\def \bea {\begin{eqnarray}}
\def \eea {\end{eqnarray}}
\def \sla {\slashed}
\usepackage{graphicx}
\begin{document}
\title{Higher time derivatives, stability and Fermi Statistics}
\author{Justo Lopez-Sarrion}
%\email[Electronic mail: ]{}
\affiliation{Departamento de F{\'i}sica, Universidad de Santiago de Chile, Casilla 307, 
Santiago, Chile.}
\author{Carlos M. Reyes}
%\email[Electronic mail: ]{}
\affiliation{Departamento de Ciencias B{\'a}sicas, Universidad del B{\'i}o B{\'i}o, Casilla 447, 
Chill\'an, Chile.}
%\date{December xxx; published as Phys.\ Rev.\ D {\bf xx}, xxxxxx (xxx)}
\begin{abstract}
We show that statistics is crucial for the instability problem derived from higher time derivatives.
In fact, and contrary to previous statements,
we check that when dealing with Fermi systems, the Hamiltonian is well bounded
and the quantum states are normalizable. Although, ghost states are still present, they do not affect unitarity under certain conditions. 
We first analyze a quantum oscillator involving Grassman variables
and then we 
generalize it to a Dirac field. 
Finally, we discuss some physical implications.
\end{abstract}
\pacs{11.10.-z, 11.10.Lm, 03.70.+k, 12.10.-g}
%\begin{keyword}
%xxxx, xxxx, xxxx, xxx
%\PACS xxx, xxx, xxx
%\end{keyword}
\maketitle
%-----------------------------------------abstract----------------------------------------------------------
In recent years there has been growing interest in field theories with higher derivatives. 
Nonlocal effects in string theory \cite{Strings}, quantum gravity \cite{Stelle} 
and noncommutative geometry \cite{NC-Geom} provide a natural
landscape for the incorporation of higher derivative terms. Also, attempts to 
incorporate such terms 
have been given in the context of dark energy \cite{D-E}, Lorentz 
violation \cite{L-V}, radiative corrections \cite{Radiative} and regularization \cite{Regularization}.

However, according to current belief, field theories with higher time
derivatives lead to an unavoidable energy instability. 
In general, this statement is a consequence of Ostrogradsky theorem \cite{OSTRO}.
A paradigmatic example is given by the well known 
Pais-Uhlenbeck model \cite{P-U} where it is possible to show explicitly
that the energy is not bounded below. One way to avoid Ostrogradsky theorem consists to 
consider a non normalizable ground state. However, this alternative not only has 
the drawbacks
of containing this non normalizable ground state, but also yields states with negative norm \cite{D-E,P-U,Ghosts}.
Although, those ghost states may cause a loss of unitarity, under certain
conditions, they do not imply a serious problem in order to define a sensible theory \cite{H-H}.
However, either the instability or the non normalizability invalids any attempts of 
a perturbative expansion.
A third approach to deal with this problem has been made by redefining the inner product by using 
$\mathcal{PT}$ symmetry \cite{Bender-Mannheim}. Others include perturbative 
iterations \cite{Jaen} and performing complex transformations \cite{Complex-Transf}. 
 
So far, nobody has explored the role played by statistics on this issue. In this work,
we propose a model somehow equivalent to the Pais-Uhlenbeck model,
but using fermion instead of boson variables. We show that our model mimics some features of the 
Pais-Uhlenbeck model: one frequency has no perturbative behavior, the phase space is increased, unitarity
may be lost when an interaction is turned on. 
However, the instability problem does not show up. 
The mechanism that allows this is essentially the same as the 
one that stabilizes the Dirac sea in the electron theory.

To begin with, let us recall the Pais-Uhlenbeck Lagrangian
\begin{eqnarray}
L_{PU}= \frac{g}{2\omega^2} \ddot q^2+\frac{1}{2} \dot q^2-\frac{1}{2}\omega^2q^2,
\end{eqnarray}
which is basically the harmonic oscillator with frequency $\omega$ plus 
a higher derivative term.
The Hamiltonian of the system after quantization can be written as 
\begin{eqnarray}
\hat H_{PU}=k_{-}\hat a^{\dag}\hat a- k_{+} \hat b^{\dag}\hat b+\frac{1}{2}(k_{-}-k_{+}), \label{pais}
\end{eqnarray}
with $k^2_{\pm}= \frac{\omega^2}{2g} (-1\mp \sqrt{1+4g})$ positive frequencies depending on $g$ and $\omega$ and 
$\hat a$, $\hat a^{\dag}$, $\hat b$, $\hat b^{\dag}$ the standard creation and
annihilation operators.  
The second term 
produces arbitrary negative energy states as can be seen 
by acting $\hat b^{\dag}$ on the empty wave function (defined by
$\hat a \Phi_0=\hat b\Phi_0=0$)
\begin{eqnarray}
\Phi_0=N \exp \left[-\frac{\sqrt{1+4g}}{2(k_{+}+k_{-})}    
(k_{+} k_{-}q^2+\dot q^2  )+\sqrt{g}q\dot q\right].
\end{eqnarray}
The fact of having energies not bounded below, invalids the perturbation formalism. Indeed,
the expectation values of some observables depending on $\dot q$,
blows up when $g$ goes to zero.
An alternative proposal would be to redefine the vacuum $\hat a \Phi_0'=\hat b^{\dag} \Phi_0'=0$,
which solves the stability problem. However, this new vacuum state is not normalizable, and negative normed states emerge.
Our proposal is to change commutators with anticommutators, namely,
to work with fermions instead of bosons.

Let us then consider the fermionic model 
\begin{eqnarray}
L_{F}= -\frac{g}{\omega}\bar\psi\ddot\psi + i\bar\psi\dot\psi -\omega\bar\psi\psi.
\end{eqnarray}
As before, this is the (fermionic) harmonic oscillator ($g=0$) plus
a higher derivative term. Without loss of generality we can assume that $\omega$ and $g$ are positive constants.
The equations of motion are,
\begin{eqnarray}
\ddot\psi= i\frac{\omega}{g}\dot\psi -\frac{\omega^2}{g}\psi,
\label{em}
\end{eqnarray}
and its conjugate. If we try a solution like $\psi= \eta_0\,e^{i\lambda t}$ we find that,
$ \lambda^2- \lambda \frac{\omega}{g} - \frac{\omega^2}{g}=0,\label{char}$
whose solutions can be expressed in terms of positive quantities $\omega_\pm$, 
\begin{eqnarray}
\omega_\pm &\equiv&\pm \lambda_\pm= \frac{\pm1+ \sqrt{1+4g}}{2g}\omega,
\label{omega}
\end{eqnarray}
where $\omega_+>\omega_->0$.

The Hamiltonian and the commutation relations which reproduce the equations of motion are,
\begin{eqnarray}
H= \frac{g}{\omega}\dot{\bar\psi}\dot\psi + \omega\bar\psi\psi,
\label{hamil}
\end{eqnarray}
and the non vanishing anticommutators
\begin{eqnarray}
\{\dot {\bar \psi},  \psi\}= i\frac{\omega}{g},
\quad
\{\dot  \psi, \bar  \psi\}= -i\frac{\omega}{g},\label{atc1}
\quad
\{\dot \psi,  \dot {\bar \psi}\}= -\frac{\omega^2}{g^2}. \label{atc2}
\end{eqnarray}
Using the Schroedinger representation in terms of Grassman variables and their derivatives, we have,
\begin{eqnarray}
\dot {\bar \psi}&=&i\frac{\omega }{g }\frac{\partial }{\partial \psi}-i\frac{\omega}{2g}\bar \psi,\label{psidot}\nonumber \\
\dot { \psi}&=&-i\frac{\omega }{g }\frac{\partial }{\partial \bar\psi}+i\frac{\omega}{2g} \psi.\label{barpsidot}
\end{eqnarray}
And the Hamiltonian in this representation is,
\begin{eqnarray}
H&=&\frac{\omega }{g}\frac{\partial }{\partial \psi}\frac{\partial }{\partial \bar \psi}+\omega\left(1+\frac{1}{4g}\right)\bar \psi\psi
\nonumber\\
&-&\frac{\omega} 
{2g}\left( \bar \psi \frac{\partial }{\partial \bar \psi} -\psi \frac{\partial }{\partial \psi}\right)-\frac{\omega}{2g}.
\label{schrhamil}
\end{eqnarray}
The first and second lines of this equation commute to each other, so it is very easy to find the eigenfunctions,
\begin{eqnarray}
\Phi_0= \frac{\sqrt 2}{(1+4g)^{\frac{1}{4}}}e^{-\frac{\sqrt{1+4g}}{2}\,\bar\psi\psi}\,,&& E_0 = -\omega_+,  \label{0}\nonumber \\
\Phi_1 = \bar\psi\quad\quad\,\quad\quad\quad\quad\quad\quad,&& E_1=\omega_- -\omega_+,\label{1}\nonumber \\
\Phi_2= \psi\quad\quad\quad\,\quad\quad\quad\quad\quad,&& E_2 = 0,\label{2}\nonumber \\
\Phi_3= \frac{\sqrt 2}{(1+4g)^{\frac{1}{4}}}\, e^{\frac{\sqrt{1+4g}}{2}\,\bar\psi\psi}\, ,&& E_3 =  \omega_{-}. \label{3}
\end{eqnarray}
Note that for $g>0$ they are ordered in increasing energy, and of course, 
the system is bounded. So, the vacuum will correspond to the lowest 
energy state, {\it i.e.}, the state $\Phi_0$. However, the  shift of 
energy respect to the lowest energy state is what physically matters. 
After this shifting we obtain $\Delta E_0=0$, 
$\Delta E_1=\omega_-$, $\Delta E_2=\omega_+$ and $\Delta E_3
= \omega_++\omega_-$ which are all positive.
Thus, we note that the energies of the first two states go 
to $0$ and to $\omega$ as $g$ goes to $0$, respectively. 
However, the last two energies are infinitely large  as $g$ tends to zero.

Let us see what happens with the normalization. If we define the Berezin measure as,
$\int d\bar\psi\,d\psi \psi\bar\psi= +1$
we can see that,
$\langle\Phi_0\vert\Phi_0\rangle =
\langle\Phi_1\vert\Phi_1\rangle=1$
which are normalizable states (if $g>-1/4$). However,
$\langle\Phi_2\vert\Phi_2\rangle=\langle\Phi_3\vert\Phi_3\rangle=-1 $
are states with negative norm. It is worth noting that, since the system 
is bounded below, in the limit of $g \to 0$, 
the ghosts states live in an arbitrary high energy level, and we should expect this sector decouples from the theory. 

Even in the case when $g$ is finite, unitarityis preserved. To see this, let us  define the new inner product, where $g$ $(\alpha\vert\beta) = 
\langle \alpha\vert \hat G\vert\beta\rangle$, where $\hat G=\exp{(i\pi \hat N_+)}$, 
being $N_+=c_+^\dag c_+$ the number of ghosts operator. This new inner product 
is the usual positive defined one. This is to say, we can write $ \langle \alpha\vert\beta\rangle=
( \alpha\vert \hat G\vert\beta)$, and then, the time evolution of amplitudes are,
\begin{eqnarray}
i\partial_t\langle \alpha\vert\beta\rangle &=& ( \alpha\vert (\hat G\hat 
H - \hat H^\dag\hat G)\vert\beta)\nonumber\\
&=&\langle \alpha\vert (H-\hat G\hat H^\dag\hat G)\vert\beta\rangle\label{probability}
\end{eqnarray}
where $\hat H^\dag$ is the usual hermitian conjugate of $\hat H$. Then, in 
order to obtain probabilities conserved we need $\hat H=\hat G\hat
 H^\dag \hat G$. In the literature, \cite {mostafa}, this is called pseudohermicity condition, and, under certain circunstances, it is possible to obtain a well defined unitary theory \cite{das}. This is the case for our modal, where $\hat G^2=I$, 
 $\hat H^\dag =\hat H$ and $[\hat G,\hat H]=0$. Then, we do not lose unitarity. 
 Adding interactions can change the situation, but there are still chances to 
 keep unitarity. In any case, when $g$ is small enough, so that the ghost sector can be considered as decouple from the theory, we can perform pertubative calculations in a safe way. This last fact, is contrary to the bosonic case, where either the instability problem or the non normalizability of ground state invalids any pertubative formalism. 

In order to clarify a little more what is going on, and 
to make a closer contact to the QFT formalism, let us consider the new variables,
\begin{eqnarray}
c_- = \alpha(i\dot\psi + \omega_+ \psi)
,&&
c_-^\dag= \alpha(-i\dot{\bar\psi} + \omega_+\bar\psi)\label{c-},
\end{eqnarray}
and,
\begin{eqnarray}
c_+= \alpha(-i\dot{\bar\psi} - \omega_-\bar\psi),&&
c_+^\dag = \alpha(i\dot\psi - \omega_-\psi)\label{c+},
\end{eqnarray}
where $\alpha=\left(\frac{g/\omega}{\omega_++\omega_-}
\right)^{\frac{1}{2}}$.
It is easy to show that the non vanishing anticommutators are,
$\{c_{-},c_{-}^\dag\}= 1, \{c_{+},c_{+}^\dag\}= -1$.
Then, the operators $c_-$, $c_-^\dag$, $c_+$ and $c_+^\dag$ are standard 
creation and annihilation operators, but $c_+$ and $c_+^\dag$ create and annihilate states of negative norm.

The original variables can be written in terms of these operators as
\begin{eqnarray}
\psi&=&\frac{c_--c^\dag_+}{\sqrt{1+4g}},\label{pisc}\nonumber \\
\dot\psi &=&-i\frac{\omega_- c_- + \omega_+ c_+^\dag}{\sqrt{1+4g}}, \label{psicdot}
\end{eqnarray}
so the Hamiltonian turns out to be 
\begin{eqnarray}
H_F = \omega_- c_-^\dag c_-  - \omega_+ c_+^\dag c_+ -\omega_+.
\end{eqnarray}
This expression can be compared with the Pais-Uhlenbeck 
Hamiltonian \ref{pais}. They are the same expressions, except for an irrelevant constant. 
However, as we see, the spectrum is bounded below, and the vacuum is 
normalizable. In fact, the true vacuum is the state annihilated by 
$c_-$ and $c_+$, namely,
\begin{eqnarray}
c_-\Phi_0&=& \alpha\frac\omega g \left(\frac{\partial}{\partial  
\bar{\psi}} +\frac{\sqrt{1+4g}}{2}\psi \right)\Phi_0 =0,\nonumber \\
c_+\Phi_0&=&\alpha\frac{\omega}{g} \left(\frac{\partial}{\partial \psi} 
-\frac{\sqrt{1+4g}}{2}\bar\psi\right)\Phi_0 =0, 
\end{eqnarray}
which agrees with the result above. Morever, the rest of the 
states are $\Phi_1=c_-^\dag\Phi_0$, $ \Phi_2=c^\dag_+\Phi_0$ and 
$\Phi_3=c^\dag_-c_+^\dag \Phi_0$ with their corresponding energies. 
Then, the $c^\dag_+$ operator is creating ghosts states in the system, 
even though the problem is bounded and hence stable.

This result can be generalized for terms with higher 
time derivatives whenever the frequencies hold real. 
Furthermore, the creation and annihilation operator formalism we have just seen,
allows to generalize straightforwardly our quantum mechanical model
to a fermionic QFT system with higher time derivatives. 

In the remaining of this work, we deal with this generalization to a field theory framework. 
The obvious covariant QFT generalization of our model is given by the Lagrangian density
\begin{eqnarray}\label{M-P-Lag}
\mathcal L=\bar \psi(i {\slashed{\partial}}-m)\, \psi-\frac{g}{\Lambda}  \bar 
\psi \,\Box\, \psi,
\end{eqnarray}
where $g$ is a dimensionless positive coupling 
constant and $\Lambda$ is an ultraviolet energy scale. This scale 
is the fingerprints of some fundamental theory and hence our model must be considered as an effective theory
valid only for energies far below that scale.

The equations of motion followed from (\ref{M-P-Lag}) are
\begin{eqnarray}
(i {\sla{\partial}}-m-\frac{g}{\Lambda}  \,\Box\,) \psi=0.
\end{eqnarray}
The solutions of the equation are given in terms of plane 
waves $\psi (x)=\int d^4 p e^{-i p\cdot x} w(p) $ where $w(p)$ is a Dirac spinor satisfying
\begin{eqnarray} \label{Eq-motion}
 \Big[ {\sla{p}}-m +
\frac{g}{\Lambda}p^2 \Big]w(p)=0.
\end{eqnarray}

Nontrivial solution of this equation imply $p^2 = m^2_\pm$
with,
\begin{eqnarray}
m_\pm = \frac{\pm 1 +\sqrt{1+4g\frac{m}{\Lambda}}}{2g}\Lambda.
\end{eqnarray}
It turns out that the $w$ spinors are the free Dirac solutions with masses $m_{\pm}$, namely
\begin{eqnarray} 
 \Big[ {\sla{p}} - m_\pm\Big] u^s_\pm (p_\pm)=0,\nonumber \\
 \Big[ {\sla{p}} + m_\pm\Big] v^s_\pm (p_\pm)=0,
 \end{eqnarray}
where the index $s$ stands for the spin quantum number, and $p_\pm = (E_\pm, {\rm p})$ with  
$ E_{\pm}=\sqrt{{\bf p}^2+m_{\pm}^2}$.
As expected, we have in the limit $g\to 0$
one regular solution $m_-\to m$ and the other $m_+$ going to infinity. 
Also, it is worth noting here that, even in the case when $m$ is zero, 
we still have a non vanishing $m_+=\Lambda/g$.

Following the same spirit as in the last part of our quantum 
mechanical model, we can define the fields,
\begin{eqnarray}
 \psi_{-} & =&\beta (i 
 {\sla{\partial}}+m_{+})\, \psi,\nonumber \\
%\psi^\dag_{-} & =&\beta (-i\dot{\psi}^\dag\gamma^0 + i\partial_i \psi^\dag\gamma^i + m_+\,\psi^\dag)
%  \\
 \psi_{+}  &=& \beta
 (i {\sla{\partial}}-m_{-})\, \psi, %\\
 %\psi_{+}^\dag & =&\beta (-i\dot{\psi}^\dag\gamma^0 + i\partial_i \psi^\dag\gamma^i - m_-\,\psi^\dag )
\end{eqnarray}
with $\beta= \left(\frac{g/\Lambda}{ m_{+}+m_{-}}\right)^{\frac{1} {2}}$.
%The Hamiltonian is,
%\begin{eqnarray}
%H =\int d^3 {\bf x}\, \left[ \frac g m \dot{\bar\psi}\dot\psi +\bar\psi\left( m +i\vec{\bf \gamma}\cdot\vec{\bf \nabla}-{\bf \nabla}^2\right)\psi \right],
%\end{eqnarray}
%and the non vanishing commutators
%\begin{eqnarray}
%\{\dot { \psi}^\dag({\bf x}),  \psi({\bf y})\}&=& i\frac{m}{g}\gamma^0\delta^3({\bf x} - {\bf y}),\\
%\{\dot  \psi({\bf x}), \psi^\dag({\bf y})\} &=&-i\frac{m}{g } \gamma^0\delta^3({\bf x} - {\bf y}),
%\\
%\{\dot \psi({\bf x}),  \dot {\psi}^\dag({\bf y}\} &=& -\frac{m^2}{g^2}\delta^3({\bf x} - {\bf y})\end{eqnarray}
The Lagrangian density in terms of these fields is
\begin{eqnarray}
\mathcal L&=&\bar \psi_{-}  (i {\sla{\partial}}-m_{-})\, 
\psi_{-}-\bar \psi_{+}  (i {\sla{\partial}}+m_{+})\, \psi_{+}.
\end{eqnarray}
%So that, the non vanishing anticommutators are,
%\begin{eqnarray}
%\{ \psi_-({\bf x}),\psi^\dag_-({\bf x})\}&=& \delta^3({\bf x}-{\bf y}),\\
%\{ \psi_+({\bf x}),\psi^\dag_+({\bf x})\}&=& -\delta^3({\bf x}-{\bf y}),
%\end{eqnarray}
Now it is easy to see that the Hamiltonian is
\begin{eqnarray}
H&=&\int d^3{\bf x}\,\left[\bar\psi_-(i\vec\nabla\cdot\vec\gamma + m_-)\psi \right.\nonumber\\
&-&\left. \bar\psi_+(i\vec\nabla\cdot\vec\gamma - m_+)\psi_+\right],
\end{eqnarray}
with the non vanishing anticommutators,
\begin{eqnarray}
\{\psi_-(\bf x),\psi^\dag_-(\bf y)\}=-\{\psi_+(\bf x),\psi^\dag_+(\bf y)\}=\delta^3(\bf x-\bf y).
\end{eqnarray}
Now, decomposing these fields in terms of plane wave solutions we find that
\begin{eqnarray}
  \psi_ {-}  &=&\sum_{s=1,2}\int \frac{d^3p }{(2\pi)^3}\frac{m_{-}}
  {E_{-}} \\
  &&\left[b_{-}^s({\bf p})e^{-ip_{-}\cdot x}u_{-}^s({\bf p})+
  d^{s\dag}_{-}({\bf p})e^{ip_{-}\cdot x}v_{-}^s({\bf p})\right],\nonumber \\
  \psi_ {+}  &=&\sum_{s=1,2}\int \frac{d^3p }{(2\pi)^3}
  \frac{m_{+}}{E_{+}} \\
  &&\left[b_{+}^{s\dag}({\bf p})e^{-ip_{+}\cdot x}u_{+}^s({\bf p})+
  d^{s}_{+}({\bf p})e^{ip_{+}\cdot x}v_{+}^s({\bf p})\right].\nonumber 
\end{eqnarray}
Notice the change respect to the standard decomposition in  $\psi_+$, 
swapping the $b$'s and $d^\dag$'s operators by $b^\dag$'s and $d$'s.
The non vanishing anticommutators are the standard ones for $b_-$ and $d_-$, that is,
 \begin{eqnarray}
\{b^s_-({\bf p}),b_-^{r\dag}({\bf k})\}&=&\delta^{sr}\delta^3({\bf p - \bf k}),\nonumber\\
\{d^s_-({\bf p}),d_-^{r\dag}({\bf k})\}&=&\delta^{sr}\delta^3({\bf p - \bf k}),
\end{eqnarray}
but for the $b_+$ and $d_+$ we have,
 \begin{eqnarray}
\{b^s_+({\bf p}),b_+^{r\dag}({\bf k})\}&=&-\delta^{sr}\delta^3({\bf p-\bf k}),\nonumber\\
\{d^s_+({\bf p}),d_+^{r\dag}({\bf k})\}&=&-\delta^{sr}\delta^3({\bf p - \bf k}).
\end{eqnarray}
The minus sign is revealing that these modes are ghosts, {\it i.e.}, 
these operators create and destroy negative norm states. 
The Hamiltonian in terms of these operators is, except for an irrelevant constant,
\begin{eqnarray}
H&=&\sum_{s}\int d^3p  \Big(E_{-}(b^{s\dag}_{-}({\bf p})  
 b_{-}^s({\bf p})+d^{s\dag}_{-}({\bf p})d^{s}_{-}({\bf p})) \nonumber \\
&&
+E_{+}(b^{s\dag}_{+}({\bf p})   b^s_{+}({\bf p})+d^{s\dag}_{+}({\bf p})d^{s}_{+}({\bf p}))\Big).
\end{eqnarray}
The vacuum is the state annihilated by all operators $b_\pm^s$ and 
$d_\pm^s$. This state is the true lowest energy state, and it is 
normalizable and well defined. Hence, again we have a stable theory, but with negative norm states above energies of order greater than $m_+\sim \Lambda$. As we discussed above, this fact ensures that if we add any known interaction, the theory would still be well 
defined, as long as the energy scale involved in the physical processes
are lower compared to the ghost mass scale.

As it was discussed for the QM model, our QFT theory will not have unitarity problems. 
Only by adding interactions they can come up. However, the loss of unitarity due to the ghost sector, 
will be negligible as we are much below the ultraviolet scale. Even more, the standard  
unitary theory is recovered when $g$ tends 
to zero. This situation is similar to the 
one pointed out in \cite{H-H}, but in our case we have a stable 
vacuum and we can make sense of the theory even without going to the Euclidean space.

Finally, to calculate propagators, it is useful to have in mind that 
the original fields are combinations of $\psi_-$ and $\psi_-$, namely,
\begin{eqnarray}
\psi(x)&=& \frac{1}{(1+4g\frac{m}{\Lambda})^{\frac{1}{4}}}\left[\psi_-(x) - \psi_+(x)\right]
%\\
%\slashed{\partial}\psi(x)&=& \frac{-i}{
%(1+4g^2)^{\frac{1}{4}}}\left[m_-\psi_-(x) + m_+ \psi_+(x)\right]
\end{eqnarray}
 And the propagator is,
\begin{eqnarray}
S(x) = \frac{1}{\sqrt{1+4g\frac{m}{\Lambda}}}\left(S_-(x) - S_+(x)\right)
\end{eqnarray}
where $S_\pm$ are the standard propagators for Dirac particles with 
masses $m_\pm$. The minus sign in front of $S_+$ is another signal of 
the ghost sector given by the $+$ modes. This propagator resembles the Pauli-Villars 
regularization, with $m_+$ as the regulator which 
render ultraviolet divergent integrals finite. However, this 
regulator works differently from the Pauli-Villars formalism in some 
situations. For instance, it produces new vertices when we introduce 
interactions when we keep gauge invariance.

Summarizing we have found that Fermi statistics can fix the problem of 
stability in higher time derivatives systems, giving a counterexample of the Ostrogradsky's theorem.
However, negative norm states are still present. These ghost sector can be controlled either by restricting ourselves to the physical sector or by keeping them far abouve the UV scale. In both cases, unitarity is preserved as long as the
relevant physical processes energy scale is far below the UV scale. And hence, it is well justified to add local interactions and  perform 
perturbative calculations in our theory. Somehow, the situation is similar 
to the standard QED, where perturbation treatment are justified only 
for energy scales far below the Landau pole. Just above this pole, the 
fine structure constant becomes negative and amplitudes like 'vacuum to 
vacuum' might be negative, leading to possible ghost states. That is a signal of the fact that QED can be considered as an effective theory valid only for certain energy sclaes. It would be 
interesting to study physical consequences derived from these new terms in relevant theories like QED-like or symmetry broken theories with fermions \cite{Parity-Breakdown}.

We thank A. Das, J. Gamboa and J.L. Cortes for useful and inspiring comments. 
J.L. acknowledges support from DICYT Grant No. 041131LS (USACH) 
and FONDECYT-Chile Grant No. 1100777. C.M.R. acknowledges 
partial support from DICYT (USACH) and Direcci\'on de Investigaci\'on de 
la Universidad del B\'{\i}o B\'{\i}o (DIUBB).

\end{document}